\documentclass[12pt]{iopart}
\pdfoutput=1
\usepackage{graphicx}
\usepackage{iopams} 
\usepackage[colorlinks,bookmarks=false,citecolor=blue,linkcolor=red,urlcolor=blue]{hyperref}
\usepackage{epstopdf}
\usepackage{color} 

\newcommand{\bc}{\begin{center}}
\newcommand{\ec}{\end{center}}
\newcommand{\be}{\begin{equation}}
\newcommand{\ee}{\end{equation}}
\newcommand{\bea}{\begin{eqnarray}}
\newcommand{\eea}{\end{eqnarray}}

\definecolor{todocolor}{rgb}{0.8,0.1,0.2}

\newcommand{\eqref}[1]{(\ref{#1})}

\begin{document}

\title{Spin-resolved entanglement spectroscopy of critical spin chains and Luttinger liquids}

\author{Nicolas Laflorencie$^{1}$ and Stephan Rachel$^2$}

\date{\today}

\address{$^1$ Laboratoire de Physique Th\' eorique, Universit\' e de Toulouse, UPS, (IRSAMC), Toulouse, France} 
\address{$^2$ Institute for Theoretical Physics, TU Dresden, 01062 Dresden, Germany} 

\begin{abstract}
Quantum critical chains are well described and understood by virtue of conformal field theory. Still the meaning of the real space entanglement spectrum -- the eigenvalues of the reduced density matrix -- of such systems remains in general elusive, even when there is an additional quantum number available such as spin or particle number.  In this paper we explore in details the properties and the structure of the reduced density matrix of critical XXZ spin-$\frac{1}{2}$ chains. We investigate the quantum/thermal correspondence between the reduced density matrix of a $T=0$ pure quantum state and the thermal density matrix of an effective entanglement Hamiltonian. Using large scale DMRG and QMC simulations, we investigate the conformal structure of the spectra, the entanglement Hamiltonian and temperature. We then introduce the notion of spin-resolved entanglement entropies which display interesting scaling features.
\end{abstract}



\section{Introduction}
Entanglement is a key concept to understand the quantum correlations at play in several condensed matter systems~\cite{calabrese_entanglement_2009}. The central object after a real space bipartition of a quantum system in a pure state $|\Psi\rangle$ is its reduced density matrix (RDM)
\be
\hat \rho_{A}= {\rm Tr}_{B} |\Psi\rangle\langle\Psi|,
\ee
where $A$ is the subsystem and ${\rm Tr}_{B}$ is the partial trace performed over the degrees of freedom of the rest of the system. Being for instance at the core of the Density Matrix Renormalization group (DMRG) algorithm~\cite{schollwock_density-matrix_2005}, entanglement has received very much attention over the past decade~\cite{osterloh_scaling_2002,vidal_entanglement_2003,calabrese_entanglement_2004,kitaev_topological_2006,stephan_shannon_2009,eisert_colloquium:_2010,casini_towards_2011,bhattacharya_thermodynamical_2013}. A quite natural quantity to study is the so-called R\'enyi Entanglement Entropy (EE) 
\be
S_q=\frac{1}{1-q}{\rm Tr}\Bigl(\hat \rho_{A}\Bigr)^q,
\ee
which in the limit $q\to 1$ yields the von Neumann EE
\be
S_1=-{\rm Tr}\Bigl(\hat \rho_{A}\ln \hat \rho_{A}\Bigr).
\ee

\begin{figure}
\centering
\includegraphics[width=0.5\columnwidth,clip]{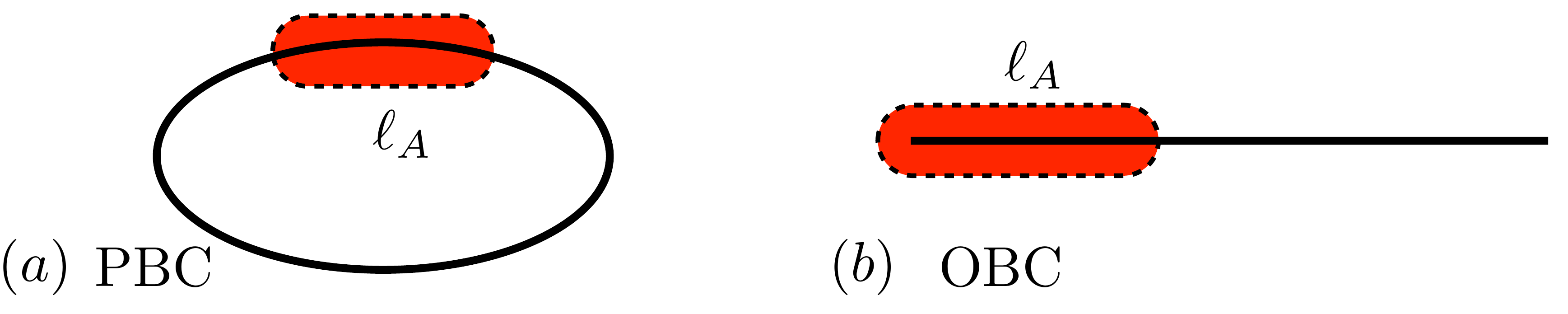}
\caption{Bipartition of the system into region $A$ (red) with length $\ell$ and the remainder $L-\ell$ where $L$ is the total length of the system. Depicted are situations for periodic boundary conditions (a) and open boundary conditions (b).}
\label{fig:OBC_PBC}
\end{figure}

R\'enyi EEs have been intensively investigated  for one dimensional systems, both analytically~\cite{calabrese_entanglement_2004,refael_entanglement_2004,calabrese-10prl095701} and numercially~\cite{vidal_entanglement_2003,laflorencie_boundary_2006}. Most importantly, a central result for a clean critical chain is its universal scaling behavior with the length $\ell$ of a subsystem~\cite{calabrese_entanglement_2009}
\be
S_q(\ell)=\frac{c}{6{\cal B}}\frac{1+q}{q}\ln \ell +~\cdots\,,
\label{eq:Sq}
\ee
where ${\cal B}=1$ for periodic boundary condition (PBC), and ${\cal B}=2$ for open boundary conditions (OBC), see Fig.~\ref{fig:OBC_PBC}. The universal character of $S_q$ appears in the prefactor $c$, which is the central charge of the underlying conformal field theory (CFT).
The dots above in Eq.~\ref{eq:Sq} correspond to subleading corrections~\cite{laflorencie_boundary_2006,cardy_unusual_2010}.

For non critical chains, EEs do not grow with $\ln \ell$ but instead saturate with the correlation length $\sim \ln \xi$~\cite{calabrese_entanglement_2004}. That is, away from a critical point the EEs approach a constant value. In the past years, also multi-interval entanglement entropy\,\cite{calabrese-09jstat11001,furukawa-09prl170602}, as well as various other measures of entanglement have been studied~\cite{eisert_single-copy_2005,song_bipartite_2012,dubail-11jsm03002,thomale-10prl116805,
greschner-13prb195101,sun-14arXiv:1408.2739,yu-14arXiv:1408.2642,gu-14arXiv:1408.2199}.

Recently, it has been suggested by Li and Haldane\,\cite{li-08prl010504} to consider not only quantities which depend on $\hat \rho_A$ like the EEs $S_q$ but instead to investigate the structure of the reduced density matrix $\hat \rho_A$ itself. They analyzed the spectrum of the reduced density matrix $\hat\rho_A$ for several fractional quantum Hall states (note that they  did not consider a real-space cut but instead an orbital cut). Writing
\begin{equation}
\hat\rho_A = \exp{\left( - \beta\mathcal{H}_E \right)}\ ,
\end{equation}
the eigenvalues $\lambda_i$ of the reduced density matrix can formally be associated with an {\it entanglement Hamiltonian} $\mathcal{H}_E$ with spectrum 
\begin{equation}
\xi_i = -\log{(\lambda_i)}\ ,
\end{equation}
dubbed {\it entanglement spectrum}. Note that in the above definition, the inverse entangle\-ment temperature $\beta=1$. Li and Haldane showed that the low-lying levels in the entanglement spectrum exhibit the same state counting as the elementary quasi-hole excitations of the fractional quantum Hall states. They further claimed that the entanglement spectrum can be used to detect {\it fingerprints} of the topological order associated with the fractional quantum Hall states\,\cite{li-08prl010504}. Soon after, Calabrese and Lefevre studied the entanglement spectrum associated with a real space cut (as shown in Fig.\,\ref{fig:OBC_PBC}) in case of a critical free-fermion chain\,\cite{calabrese-08pra032329}. As one of their central results, they found the distribution of the entanglement levels. They further pointed out that the generalization to other critical chains involves the parameters of the corresponding CFT. More recently, a correspondence between the low-lying part of the ES and the energy spectrum of a boundary CFT has been proposed giving the opportunity to extract the boson compactification radius directly from the ES\,\cite{lauchli_operator_2013}.

For gapped (spin) chains, the investigation of the ES has led to the discovery that in certain phases {\it all} entanglement levels are two-fold degenerate\,\cite{pollmann-10prb064439}. In the meantime this degeneracy has been interpreted as one of the hallmarks of symmetry protected topological (SPT) phases\,\cite{pollmann-11prb075102,chen-12s1604}. In topologically trivial phases, this degeneracy is absent. Of course, there are natural degneracies in the ES when a conserved U(1) current is present such as particle number or $z$-compoment of spin $S^z$. For instance, if the Hamiltonian commutes with $S^z$ of the total system, then the reduced density matrix $\hat\rho_A$ and the spin operator $ S_A^z$ of subsystem $A$ must commute as well,
\begin{equation}
[  S_A^z , \hat \rho_A ] = 0\ .
\end{equation}
As a consequence, each entanglement level $\xi_1$ associated with a finite $S_A^z\not= 0$ must have a partner $\xi_2 = \xi_1$ associated with $-S_A^z$ (provided the spin inversion symmetry is preserved by the Hamiltonian). If even the full SU(2) symmetry is preserved by the Hamiltonian, then the SU(2) multiplet structure is also present in the ES. The degeneracies due to spin multiplet structure (or, similarly, particle conservation) are present in both critical and gapped systems.

The presence of a conserved U(1) current also links to another interesting quantity. The variance or fluctuations of spin or particle number, respectively, defined for a bipartite subsystem $A$ behaves itself as a measure of entanglement\,\cite{song_bipartite_2012}. Recently, it has been pointed out that the EEs and these bipartite fluctuations share various properties\,\cite{song_general_2010,song_bipartite_2012,rachel-12prl116401}. In case of free fermions, exact relations between the Renyi EEs and the full set of charge cumulants have been established\,\cite{klich-09prl100502,song-11prb161408,song_bipartite_2012}. In certain cases, bipartite fluctuations can be even used to {\it measure} the entanglement spectrum of quantum Hall states\,\cite{petrescu-14arXiv:1405.7816}.

Nonetheless, the general knowledge of the entanglement spectrum and its implications in critical chains are rather limited, even when an additional quantum number such as spin or particle number is available. In order to shed some more light on this quantity, we will in the following investigate the entanglement spectrum of the XXZ spin chain as a paradigm of critical chains with conserved $S^z$ quantum number. We aim to analyze the difference subspaces associated with different values of $S^z$. We will also consider entanglement entropies which are restricted to a fixed $S^z$, dubbed spin-resolved EEs, and study their scaling behavior.

The paper is organized as follows. In Section 2, we will analyze the reduced density matrix and the entanglement Hamiltonian of the XXZ spin chain in detail, compare with predictions form the literature, and eventually consider the spin-resolved density matrix. Our findings are substantiated with large scale DMRG and QMC simulations. In Section 3, we elaborate further on the CFT-related properties of the ES. Then we discuss spin resolved entanglement entropies, and conclude in Sec.\ 4.


\section{Reduced density matrix and entanglement Hamiltonian for critical XXZ chains}
We start from the one dimensional $S=\frac{1}{2}$ XXZ model, governed by the following Hamiltonian
\be
{\cal{H}}_{\rm xxz}=\sum_{i=1}^{L+1-{\cal B}}\left(S_{i}^{x}S_{i+1}^{x}+S_{i}^{y}S_{i+1}^{y}+\Delta S_{i}^{z}S_{i+1}^{z}\right),
\label{eq:XXZ}
\ee
where ${\cal B}=1$ or $2$ accounts for boundary conditions, as defined in Eq.~\eqref{eq:Sq}. This model displays critical correlations for $-1<\Delta\le 1$ with a continuously varying Luttinger liquid parameter 
\be
K_{\rm LL}=\frac{1}{2\arccos(-\Delta)/\pi}\ .
\label{eq:K}
\ee
Following the setups sketched in Fig.~\ref{fig:OBC_PBC}, when a critical XXZ chain is cut in two parts ($A$ of length $\ell$ and the rest), the leading term of R\'enyi entanglement entropies is given by Eq.~\eqref{eq:Sq}, with a central charge $c=1$, as verified numerically for instance in Refs.~\cite{vidal_entanglement_2003,laflorencie_boundary_2006,fuehringer-08adp922,rachel-09prb180420}.  
While the knowledge of $S_q$ is fully relevant for the characterization of entanglement properties of a given system, it does not reveal the  complexity of the reduced density matrix itself, and it is expected that much more information can be obtained from the eigenvalues of the reduced density matrix: the entanglement spectrum.

\subsection{Block diagonal RDM and distribution of eigenvalues}
\subsubsection{Eigenvalues distribution: DMRG results and finite size effects}
In the following, we show and analyze numerical data for the XXZ spin chain which was obtained within the density matrix renormalization group (DMRG)\,\cite{white92prl2863,schollwock_density-matrix_2005}. Apart from the fact that the DMRG method is probably the most powerful technique for one-dimensional quantum systems at zero temperature, it features another major advantage for the study of quantum entanglement in many-body systems: the reduced density matrix is the key quantity (and not the wavefunction or the Green's function) which is calculated and optimized permanently. That is, whenever one applies the DMRG method to a problem, its entanglement entropies and entanglement spectrum is immediately known--without any extra computational costs. All DMRG results presented in this paper are computed with OBC and for system sizes ranging from $L=100$ to $L=2000$ lattice sites. We always performed 10 DRMG sweeps and kept the discarded entropy below $10^{-14}$.

\begin{figure}[h!]
\centering
\includegraphics[width=0.72\columnwidth,clip]{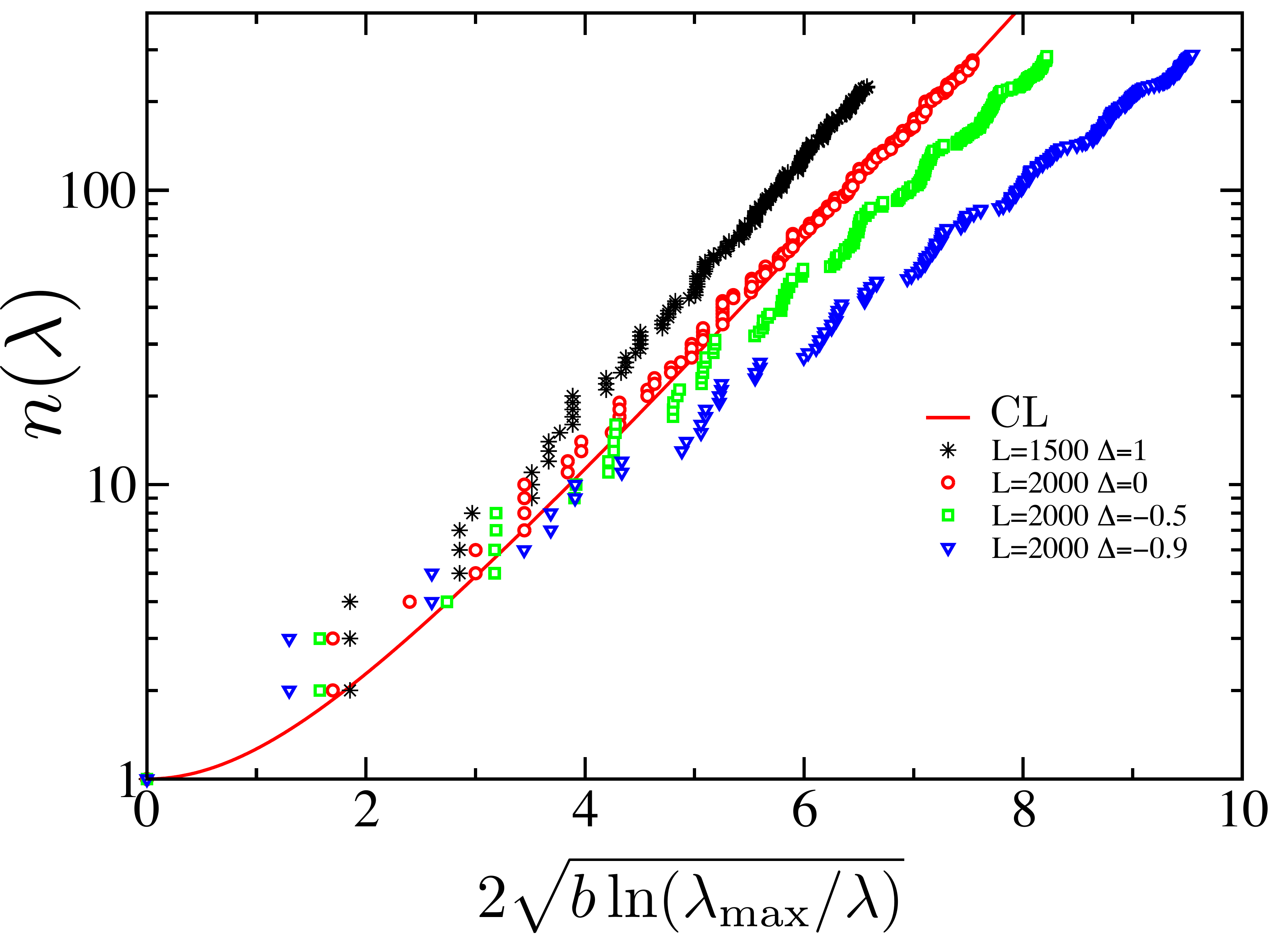}
\caption{Eigenvalues distribution $n(\lambda)$ obtained form DMRG for various anisotropies $\Delta$ for system sizes $L\ge 1500$ with OBC and partitions at $L/2$. The CL expression is from Calabrese-Lefevre~\cite{calabrese-08pra032329} Eq.~\eqref{eq:nlambda}.}
\label{fig:distrib}
\end{figure}

We expect from Calabrese and Lefevre (CL)~\cite{calabrese-08pra032329} the mean number of eigenvalues larger than a given $\lambda$ to be
\be
n(\lambda)=I_0\Bigl(b\ln(\lambda_{\rm max}/\lambda)\Bigr),
\label{eq:nlambda}
\ee
where $I_0$ is the modified Bessel function of first kind, $\lambda_{\rm max}$ the largest eigenvalue, and $b=-\ln\lambda_{\rm max}$.
DMRG results are shown in Fig.~\ref{fig:distrib} where it is very interesting to notice that the CL formula works remarkably well for the XX point (corresponding to free fermions) but we observe some significant deviations for interacting cases $\Delta\neq 0$. Such deviations have already been obseved, for instance in Refs.~\cite{pollmann-09prl255701,alba_entanglement_2012}. In particular, for attractive $\Delta<0$, $n(\lambda)$ underestimates the analytical prediction~\cite{alba_entanglement_2012}. Conversely for repulsive interaction $\Delta>0$, $n(\lambda)$ overestimates the CL-curve. While we are dealing with already very large systems (up to $L=2000$ sites), it is still possible that we are facing finite size effects which apparently change sign with the sign of anisotropy and are much smaller for $\Delta=0$. 

\begin{figure}[t!]
\centering
\includegraphics[width=.7\columnwidth,clip]{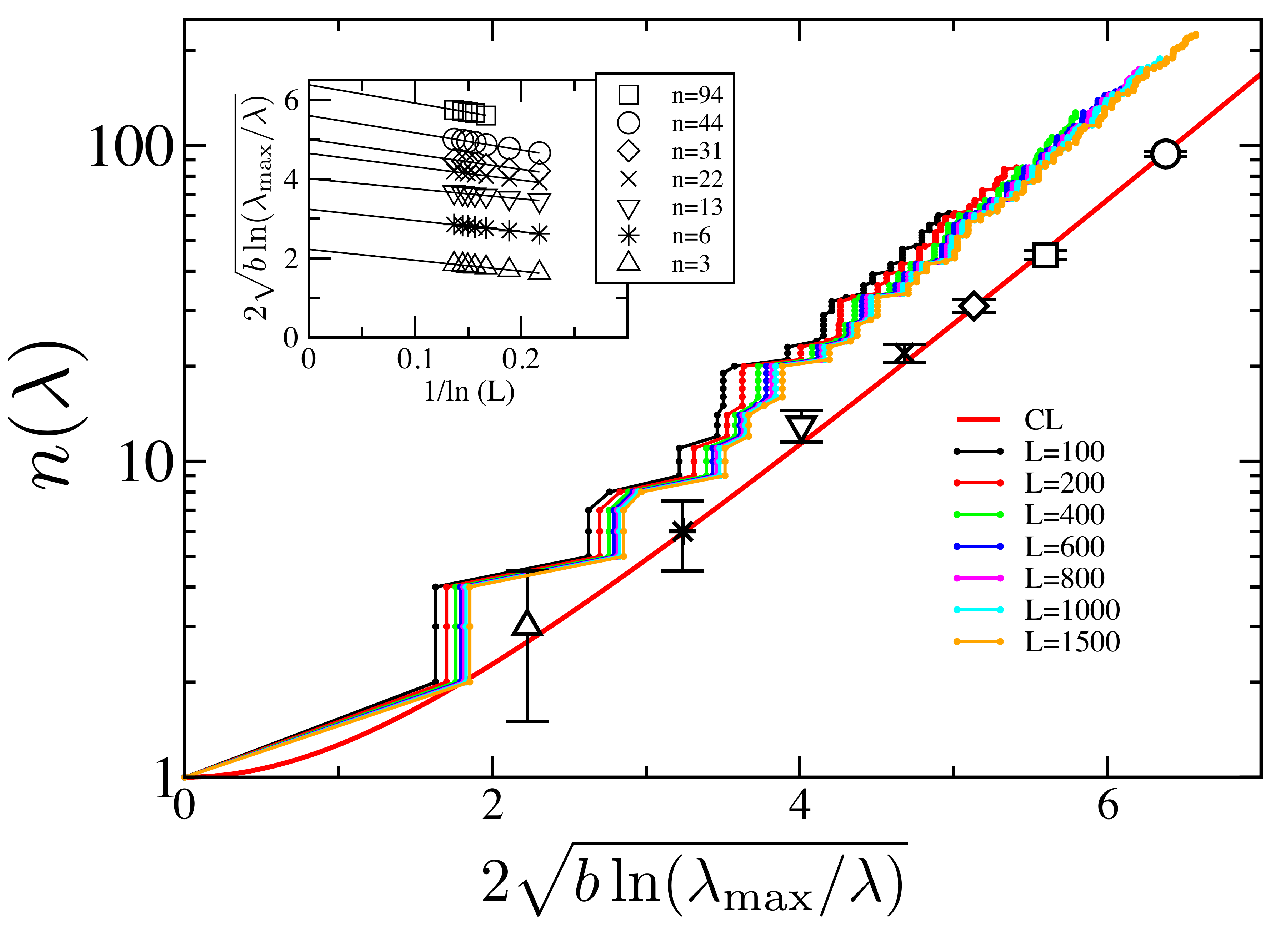}
\caption{Finite size convergence of $n(\lambda)$ towards the Calabrese-Lefevre formula Eq.~\eqref{eq:nlambda}. DMRG data for $\Delta=1$. Error bars reflect the uncertainty due to the vertical jumps in the $n(\lambda)$ curves. Inset: logarithmic convergence to the thermodynamic limit where the lines are linear fits.}
\label{fig:distrib_fse_delta1}
\end{figure}

Strong finite size effects are indeed responsible for the observed deviation, as displayed in Fig.~\ref{fig:distrib_fse_delta1} for the SU(2) Heisenberg point $\Delta=1$. There, $n(\lambda)$ are plotted for all available sizes $L=100,\cdots, 1500$, and infinite size extrapolations are performed for 7 values of $n$. As shown in the inset of Fig.~\ref{fig:distrib_fse_delta1}, the convergence to the thermodynamic limit is logarithmically slow $\sim 1/\ln L$. Nevertheless, the CL expression (red curve) gives a very good description of the DMRG data, once the thermodynamic limit is taken.
We have repeated the same analysis for $\Delta=-0.5$ and $-0.9$ (data not shown here) and also we found numerically a logarithmic convergence to the CL expression, but with an opposite sign. Our data suggest that the prefactor of the $1/\ln L$ correction has the sign of $\Delta$, and as we see in Fig.~\ref{fig:distrib}, vanish at the free-fermion point $\Delta=0$. An analytical understanding of such finite size effects is needed and certainly calls for further works.

\subsubsection{Spin-resolved RDM}
We now turn to the internal structure of the RDM of the XXZ chain, which is block diagonal, each block corresponding to the subsystem magnetization $S_A^z=0,\pm 1,\pm 2,\cdots,\pm \ell/2$. Therefore one can diagonalize separately each sector $S^z_A=\pm m$. The spin inversion symmetry yields that only $m\ge 0$ can be considered. The size ${\cal{D}}(m,\ell)$ of each $m$-sector is exponentially large with $\ell$ for finite $m$:
\bea
{\cal{D}}(m,\ell)&=&\frac{\ell!}{(\frac{\ell}{2}-m)!(\frac{\ell}{2}+m)!}\nonumber\\
&\simeq & 2^\ell\sqrt{\frac{2}{\pi\ell}}\exp(-\frac{4m^2}{\ell}),~~~(m\ll\ell).
\label{eq:Dm}
\eea
Nevertheless, the power of DMRG allows to access ground-state properties and in particular for our present purpose entanglement estimates with a very high precision by keeping only a very small number of eigenstates of the RDM, as compared to the exponentially large ${\cal{D}}(m,\ell)$ (see~\ref{sec:appendix}).
\begin{figure}[t]
\centering
\includegraphics[width=.85\columnwidth,clip]{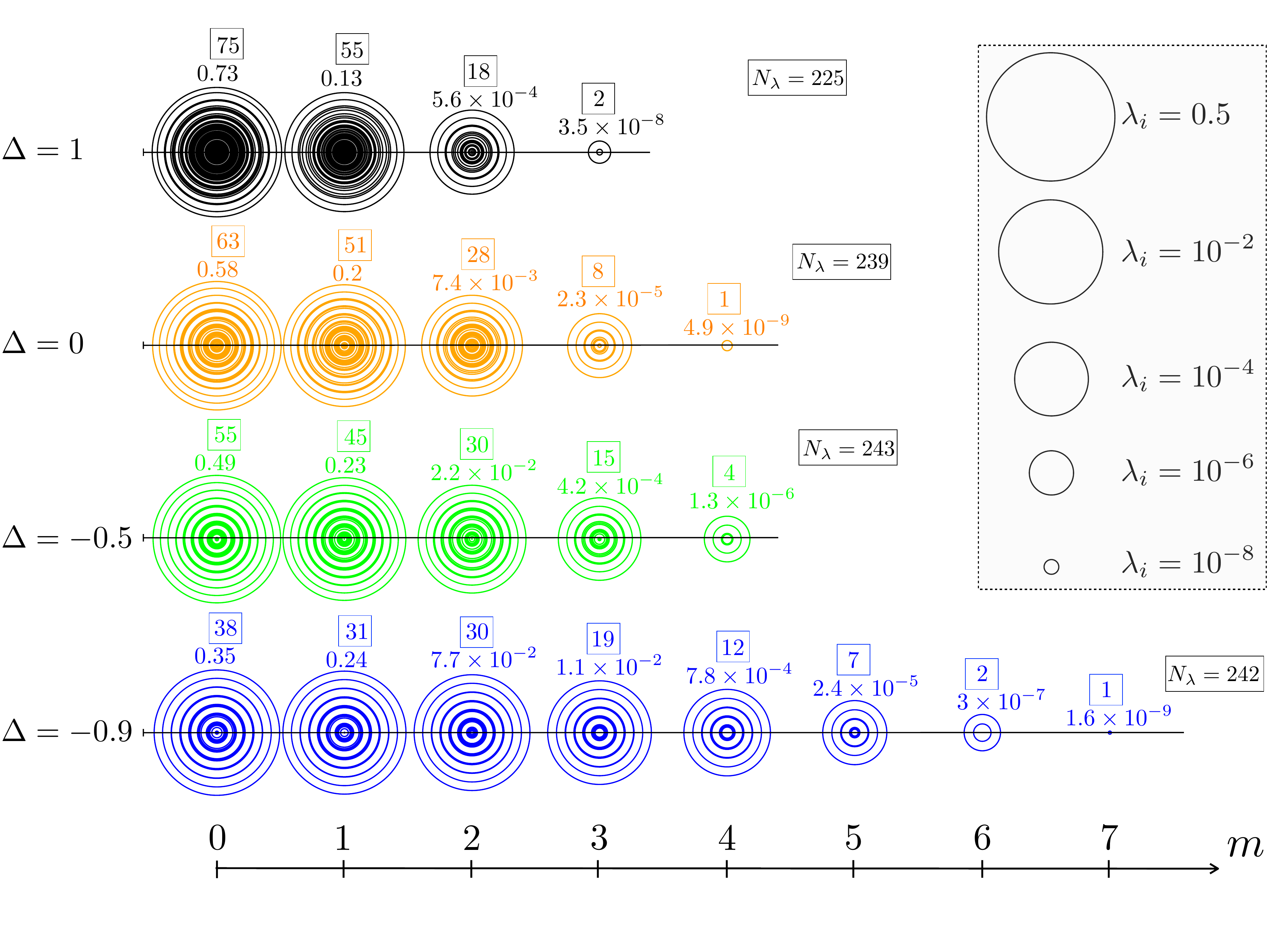}
\caption{Weights of the eigenvalues of the RDM $\hat \rho_A$ shown for 4 different values of the Ising anisotropy $\Delta$. DMRG results for $L=1500$ sites with OBC are displayed as a map showing the individual eigenvalues $\lambda_{i}$ for each sector $S^z_A=\pm m$ in a logarithmic scale (see legend). Only eigenvalues larger than $\lambda_{\rm min}=10^{-9}$ have been kept, $N_\lambda$ being the total number of such states. For each couple $(m,\Delta)$ we show above the circles (whose radii materialize the $\lambda_i$) the number of states in this subsector, and the total weight $p_m$ (see text).}
\label{fig:RDM}
\end{figure}

In Fig.~\ref{fig:RDM} we present a map of the eigenvalues of the RDM for four representative values of the Ising anisotropy $\Delta=1,~0,~-0.5,~-0.9,$ for chains of length $L=1500$ with OBC. Here, only states with weight $\lambda>10^{-9}$ were retained\,\footnote{There is no physical or numerical reason for it; merely for practical reasons we omit entanglement levels $\lambda<10^{-9}$ as this is beyond any accuracy within this analysis.}. Interestingly, while the number of kept state $N_\lambda$ does not vary so much with $\Delta$, the spin-resolved structure turns out to be qualitatively different across the critical regime. Indeed, at the Heisenberg point $33\%$ of the states lie in the $m=0$ sector with a weight $0.73$, and there is no left state for $m>3$. Conversely, close to the ferromagnetic point at $\Delta=-0.9$, the $m=0$ sector represents a total weight of 0.35 with only $15.7\%$ of the states, but one finds states up to $m=7$. Note that if the distribution of the eigenvalues was uniform, from Eq.~\eqref{eq:Dm} one would get for $L=1500$ sites a weight of only $2\%$ for the $m=0$ sector and $1.8\%$ for $m=7$.

One one hand, the result shown in Fig.~\ref{fig:RDM} is no so surprising since the magnetic correlations of the spin chain $\langle S_i^z S_{i+r}^z\rangle$ are dominated by antiferromagnetic quasi-order $\sim (-1)^r/r^{2K_{\rm{LL}}}$ for $\Delta>0$ whereas the ferromagnetic component $\sim 1/r^2$ dominates the other regime $\Delta<0$. One the other hand, as far as entanglement properties are concerned, we expect universality for both entanglement entropies Eq.~\eqref{eq:Sq} {\it{and}} spectra Eq.~\eqref{eq:nlambda} across the full critical regime. Nevertheless, microscopic details of the RDM, in particular the spin-resolved structure, appears to be a key feature that we now study in detail.

\subsection{Entanglement Hamiltonian and entanglement temperature}
\subsubsection{Quantum/Thermal mapping}
In order to get a better understanding of the RDM structure, it is very instructive to investigate the entanglement Hamiltonian. Interestingly, it was recently argued~\cite{lauchli_operator_2013} that the entanglement spectrum of ${\hat{\rho}}_A$ can be directly related to the energy spectrum of an open XXZ chain. 
We therefore expect the RDM of subsystem $A$ to be written as the following thermal density matrix 
\be
{\hat{\rho}}_A=\exp\left({-\beta_{\rm ent.}{\cal{{{H}}_{\rm xxz}^{\rm obc}}}}\right),
\label{eq:rhoA}
\ee
where ${\cal{{H}}}_{\rm xxz}^{\rm obc}$ is the Hamiltonian of an open XXZ chain Eq.~\eqref{eq:XXZ} (${\cal B}=2$) with an energy shift such that the free energy $\propto \ln Z=0$.
The inverse entanglement temperature $\beta_{\rm ent.}=1/T_{\rm ent.}$ can be determined directly from the fact that the entanglement entropy of subsystem $A$ has to match exactly the thermal entropy of the effective Hamiltonian at a temperature $T_{\rm ent.}$, this remaining true for any R\'enyi order $q$. 

In the low temperature regime $1\gg T/u\gg 1/\ell$, the extensive part of the thermal R\'enyi entropies of an XXZ chain of length $\ell$ (boundary conditions do not change this leading behavior) is~\cite{luitz_improving_2014}
\be
S^{\rm th}_q=\frac{\pi c}{6u}\left(1+\frac{1}{q}\right)\ell T.
\ee
When identified with $S_q(\ell)$ Eq.~\eqref{eq:Sq}, it yields for the entanglement temperature
\be
T_{\rm ent.}=\frac{u\ln(\ell/\ell_0)}{{\cal B}\pi\ell},
\label{eq:Teff}
\ee
with ${\cal B}=1$ (resp. ${\cal B}=2$) for PBC (resp. OBC). Note that one could also get this result from the bipartite fluctuation of magnetization~\cite{song_general_2010,song_bipartite_2012} 
\be
C_2(\ell)=K_{\rm LL}/({\cal B}\pi^2)\ln(\ell/\ell_0),
\label{eq:C2}
\ee
 which, in the thermal ensemble, is simply the Curie constant of the entanglement Hamiltonian $\chi T = (K_{\rm LL}\ell T) /(u\pi)$, equally leading to the same entanglement temperature Eq.~\eqref{eq:Teff}. Note that the above scalings are just the leading part, ignoring subdominant terms.

%
%
\subsubsection{Partition function of the entanglement Hamiltonian}

The Hamiltomian of an open critical XXZ chain, when irrelevant operators are ignored, is equivalent to a free boson model whose partition function at inverse temperature $\beta$ is known~\cite{eggert_magnetic_1992,sirker_thermodynamics_2008}:
\be
Z(\ell,\beta)= {\zeta}(\ell,\beta)\sum_{m=-\frac{\ell}{2}}^{\frac{\ell}{2}}\exp\left(-\beta\frac{\pi u}{2K_{\rm{LL}} \ell}m^2\right),
\label{eq:Z}
\ee
where ${\zeta}(\ell,\beta)=\prod_{n=1}^{\infty}\left[2\sinh\left(\frac{u\pi}{4\ell T}n\right)\right]^{-1}$.
From this expression, we immediately see that the weights $p_m$ of the sectors having $S^z=\pm m$ have a gaussian distribution
with a variance $\sigma^2=(K_{\rm{LL}}\ell T) /(u\pi)$. Therefore if the quantum/thermal correspondence is quantitatively correct, we expect the spin-resolved weights of the RDM $p_m=\sum\lambda_i^{(m)}$ (where $\lambda_i^{(m)}$ are the eigenvalues of the RDM in a given sector $m$) to be described by the gaussian distribution
\be
p_m(\ell)=\frac{1}{\sqrt{2\pi \sigma^2}}\exp(-\frac{m^2}{2\sigma^2}),
\label{eq:pm}
\ee
with a variance $\sigma^2=\sum_m m^2 p_m-(\sum_m m p_m)^2=C_2(\ell)$ which is nothing but the bipartite fluctuation of magnetization Eq.~\eqref{eq:C2} for which the scaling is well-known~\cite{song_general_2010,song_bipartite_2012}, as displayed in Eq.~\eqref{eq:C2}. This leads to 
\be
p_m(\ell)=\sqrt\frac{{\cal B}\pi}{2K_{\rm{LL}}\ln(\frac{\ell}{\ell_0})}\exp\left(-\frac{{\cal B}\pi^2 m^2}{2K_{\rm{LL}}\ln(\frac{\ell}{\ell_0})}\right).
\label{eq:pmC}
\ee
Interestingly, one can compare the above expression for $p_m(\ell)$ with the relative size of each subsector $m$ given from Eq.~\eqref{eq:Dm} by ${\cal{D}}_m(\ell)/2^\ell\simeq\sqrt{\frac{2}{\pi\ell}}\exp(-\frac{4m^2}{\ell})$. Both display gaussian distributions, but with quite different variances.

\begin{figure}[h]
\centering
\includegraphics[width=.8\columnwidth,clip]{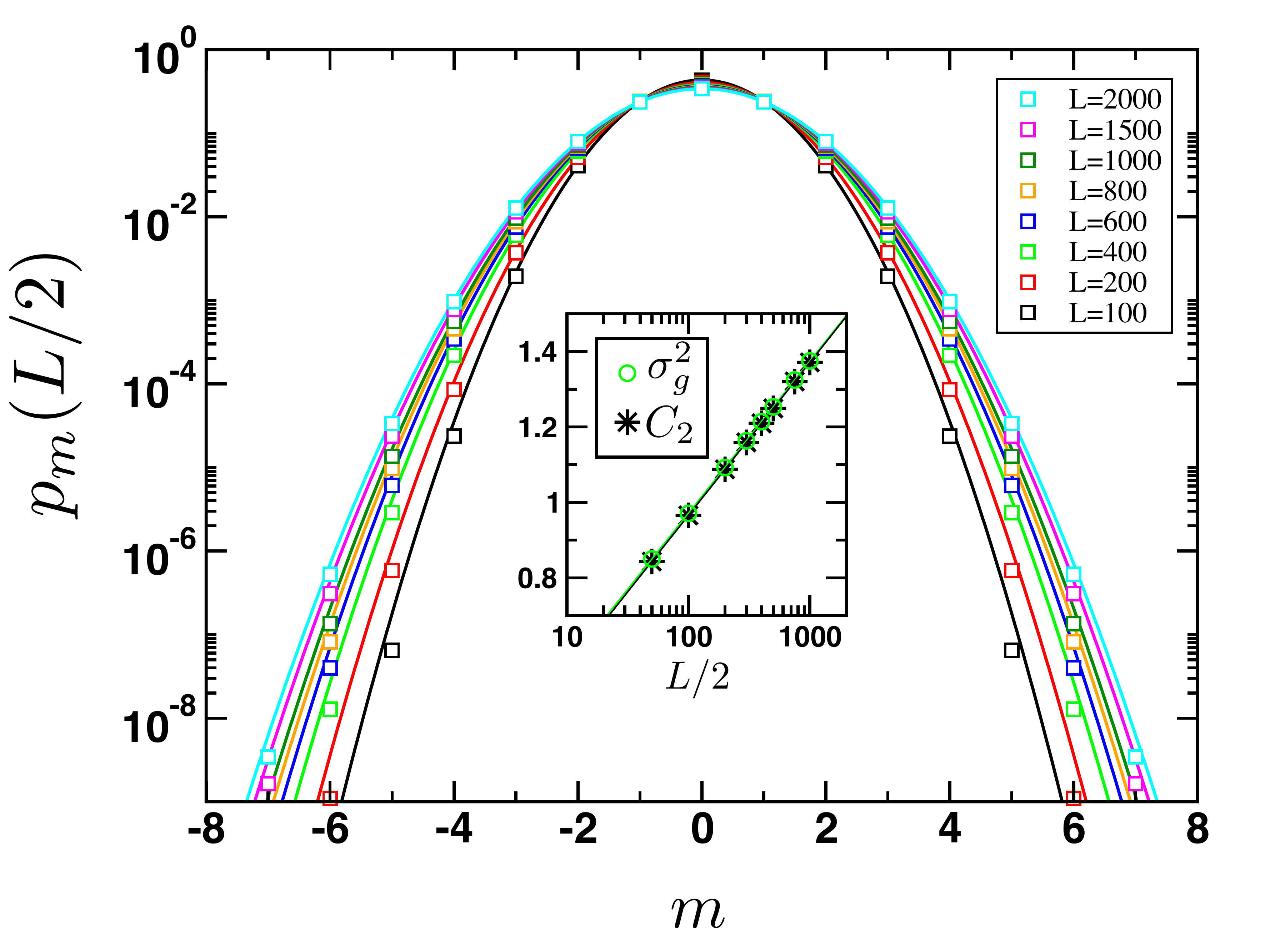}
\caption{DMRG results for the weights $p_m(L/2)$ per sector of the RDM for the XXZ chain at $\Delta=-0.9$ and OBC, shown for various chain lengths $L$ with a bipartition [Fig.~\ref{fig:OBC_PBC} (b)] at $L/2$. Full lines are fits to the gaussian form Eq.~\eqref{eq:pm} with a gaussian variance $\sigma_g^2$ displayed in the inset (green circles) where the second cumulant $C_2(L/2)$ is also shown (black stars). Both are fitted to the form Eq.~\eqref{eq:C2} with ${\cal B}=2$, $\ell_0=0.416$ and $K_{\rm LL}=3.476$ (black line) for $C_2$ and $\ell_0=0.391$ and $K_{\rm LL}=3.458$ (green line) for $\sigma_g^2$. Note that data for $L=1500$ are the same as the map in Fig.~\ref{fig:RDM}.}
\label{fig:pmdmrg}
\end{figure}

\subsection{Numerical results}
\subsubsection{DMRG}

We find that DMRG data are in very good agreement with the predicted gaussian distribution for $p_m$ Eq.~\eqref{eq:pm}, as displayed in Fig.~\ref{fig:pmdmrg} where $p_m(\ell=L/2)$ has been computed for the XXZ chain at $\Delta=-0.9$ (OBCs imposed) for various chain lengths $L=100,\ldots, 2000$. Gaussian fits yield a gaussian variance $\sigma^2_g$ which agrees perfectly with the second cumulant $C_2=\sum_mm^2 p_m$, as shown in the inset of Fig.~\ref{fig:pmdmrg}. There, both quantities are fitted to the logarithmic scaling Eq.~\eqref{eq:C2} with ${\cal B}=2$, giving for $C_2$ ($\sigma^2_g$) $\ell_0=0.416$ ($0.391$) and a Luttinger parameter $K_{\rm LL}=3.476$ ($3.458$), which compares very well to the exact value $K_{\rm LL}=3.4827$ from Eq.~\eqref{eq:K}. We clearly note that the agreement with the gaussian distribution is better for increasing system sizes. This is not surprising since DMRG is performed here for OBC ({\it cf.} Fig.~\ref{fig:OBC_PBC}(b)) where boundary effects are known to introduce finite size corrections to the leading scaling behavior~\cite{laflorencie_boundary_2006}. Furthermore, irrelevant terms have been ignored in the partition function Eq.~\eqref{eq:Z}~\cite{sirker_thermodynamics_2008}.

\begin{figure}[t!]
\centering
\includegraphics[width=.85\columnwidth,clip]{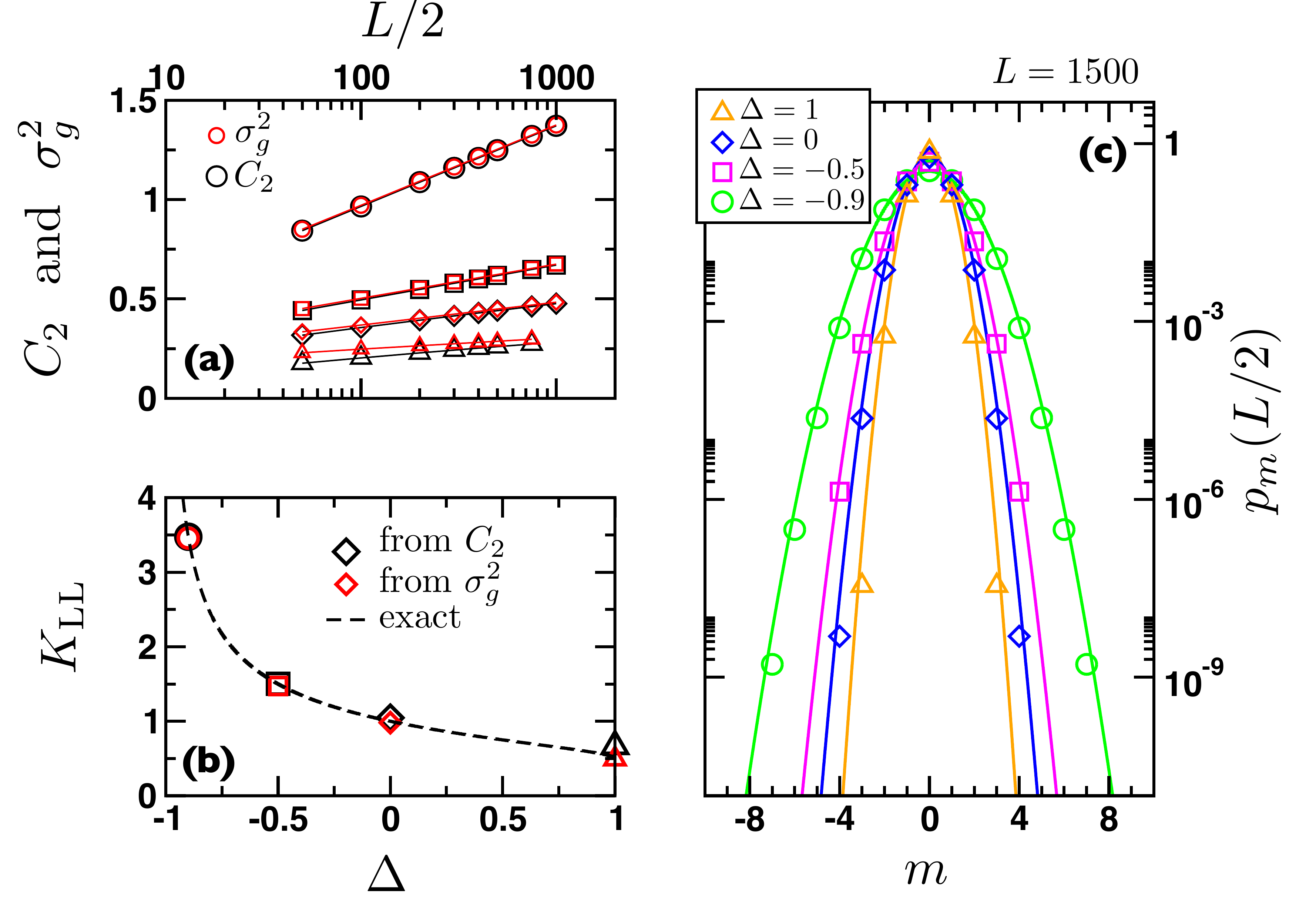}
\caption{DMRG results for the weights $p_m(L/2)$ per sector $m$ of the RDM for the XXZ chain with OBC at $\Delta=-0.9, -0.5, 0, 1$ shown in panel (c) for $L=1500$ sites. Full lines are fits to the gaussian form Eq.~\eqref{eq:pm} with a gaussian variance $\sigma_g^2$ displayed in the panel (a) (red symbols) together with the second cumulant $C_2$ (black symbols). Panel (b) shows the Luttinger liquid exponent $K_{\rm LL}$  extracted either from the second cumulant Eq.~\eqref{eq:C2} (black diamonds) or from a gaussian variance assuming Eq.~\eqref{eq:pm} (red diamonds), both being compared to the exact expression Eq.~\eqref{eq:K}. Note again that data are the same as the map in Fig.~\ref{fig:RDM}}
\label{fig:pmdmrg2}
\end{figure}

The same analysis can be repeated for other values of the anisotropy $\Delta$, as shown in Fig.~\ref{fig:pmdmrg2}. The gaussian distribution Eq.~\eqref{eq:pm} correctly describes $p_m$, as displayed in Fig.~\ref{fig:pmdmrg2} (c) for $L=1500$ and $\Delta=-0.9,-0.5,0,1$. Again one can extract both the second cumulant $C_2$ and the variance $\sigma_g^2$ obtained from a fit to the gaussian form Eq.~\eqref{eq:pm}. This is plotted in  Fig.~\ref{fig:pmdmrg2} (a) where both quantities scale with $\ln L$. Note, however, that the almost perfect agreement observed for $\Delta=-0.9$ becomes gradually less good when $\Delta$ increases, as expected since irrelevant and boundary corrections increase~\cite{sirker_chain_2007,sirker_thermodynamics_2008}. Nevertheless, the prefactor of the log behavior can be extracted, and is plotted in Fig.~\ref{fig:pmdmrg2} (b) {\it{vs.}} $\Delta$ where it compares quite well to the exact expression for the Luttinger exponent Eq.~\eqref{eq:K}.

This validates the open XXZ chain as the correct entanglement Hamitonian with an entanglement temperature given by Eq.~\eqref{eq:Teff}. One can also compare DMRG with quantum Monte Carlo computations both at zero and finite temperature, as we do now.

\subsubsection{Quantum Monte Carlo approach}
A similar study can be done using quantum Monte Carlo (QMC) simulations. While a direct access to the individual eigenvalues $\lambda_i^{(m)}$ of the RDM is practically out of reach within QMC~\cite{chung_entanglement_2014,luitz_improving_2014}, one can nevertheless sample very efficiently the diagonal of the reduced density matrix, as recently introduced in a serie of papers~\cite{luitz_universal_2014,luitz_shannon-renyi_2014,luitz_participation_2014}. The RDM being block diagonal with respect to $m$, one can also compute with QMC the trace for each sector, and therefore access $p_m$. Contrary to DMRG, PBC do not introduce additional computational costs to the QMC calculations. Moreover, one can also access finite temperature physics, while DMRG is most efficient for $T=0$ ground-state properties (although efficient finite-$T$ DMRG algorithms are available). We will exploit finite-$T$ QMC calculations below when comparing directly the distributions $p_m$ at $T=0$ for bipartite systems to finite-$T$ $p_m$ for the entanglement Hamiltonian.

\begin{figure}[t!]
\centering
\includegraphics[width=.75\columnwidth,clip]{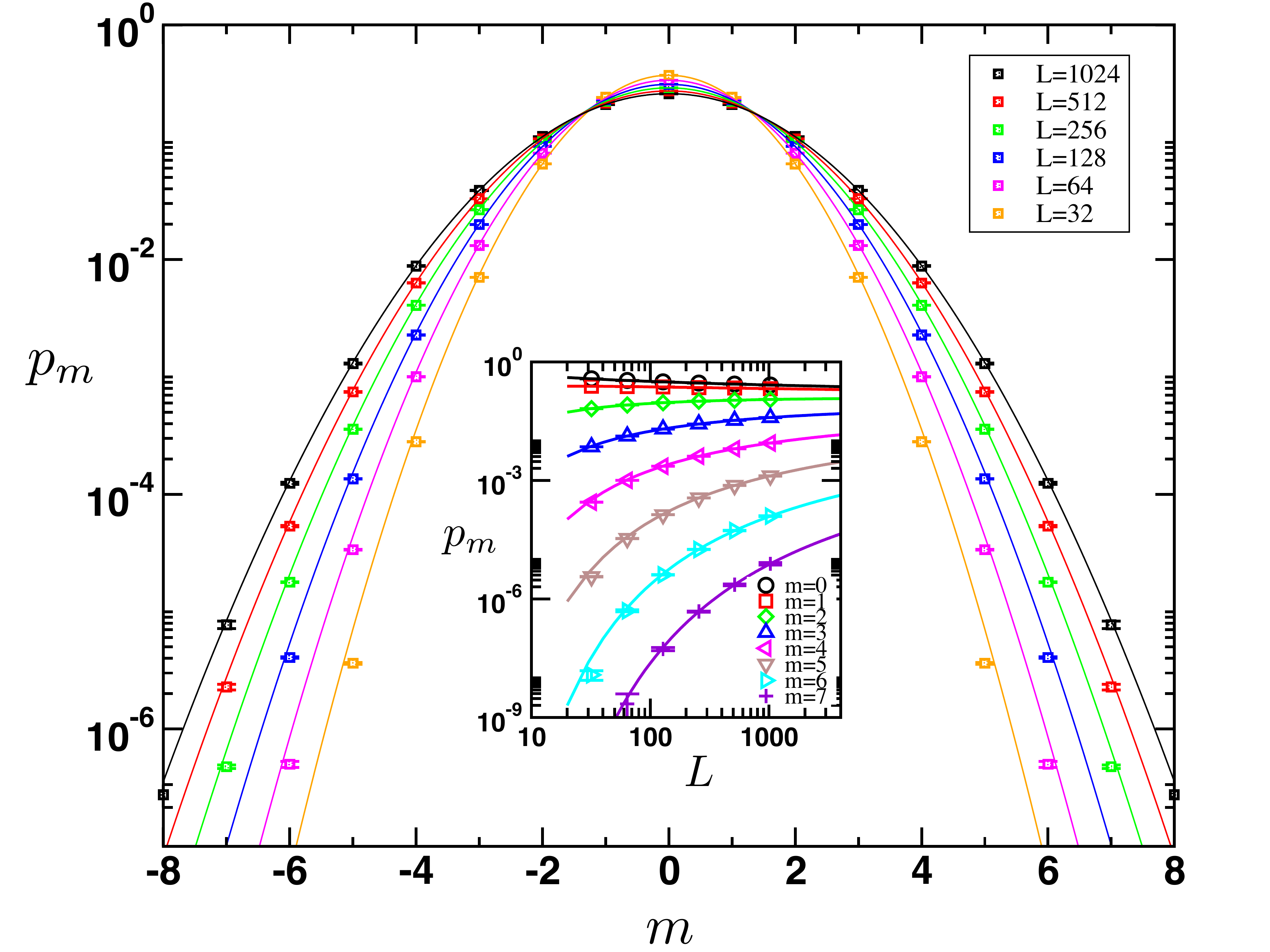}
\caption{Zero temperature QMC results for the weights $p_m(L/2)$ per sector of the RDM for the XXZ chain at $\Delta=-0.9$ and PBC, shown for various chain lengths $L$ with a bipartition [Fig.~\ref{fig:OBC_PBC} (b)] at $L/2$. Full lines are Eq.~\eqref{eq:pmC} with ${\cal B}=1$ and $K_{\rm{LL}}$ and $\ell_0$ are fit parameters displayed below in Table~\ref{tab:1}.}
\label{fig:qmc1}
\end{figure}

\paragraph{Bipartition at zero temperature---}
We present first $T=0$ QMC results for PBC (setup (a) in Fig~\ref{fig:OBC_PBC}) with a bipartition at $\ell=L/2$ for the same anisotropy $\Delta=-0.9$ as previously analysed for DMRG data. As shown in Fig.~\ref{fig:qmc1}, the gaussian behavior Eq.~\eqref{eq:pmC} is again nicely reproduced with a variance perfectly described by  $C_2(\ell)=(K_{\rm{LL}}/\pi^2)\ln(\ell/\ell_0)$. Fits are performed either {\it{vs.}} $m$ for fixed length $L$ (main panel of Fig.~\ref{fig:qmc1} and left in Tab.~\ref{tab:1}) or {\it{vs.}} $L$ for fixed magnetization $m$ (inset of Fig.~\ref{fig:qmc1} and right in Tab.~\ref{tab:1}). The agreement with the exact value of the Luttinger liquid parameter $K_{\rm LL}$ is again excellent.

\begin{table}[h!]
\bc
\begin{tabular}{c|c|c}
$L$&$K_{\rm LL}$&$\ell_0$\\
\hline
$32 $ & $3.511 $ & $0.63 $\\
$ 64$ & $ 3.519$ & $ 0.64$\\
$ 128$ & $ 3.485$ & $0.61 $\\
$ 256$ & $3.479 $ & $ 0.61$\\
$ 512$ & $3.48 $ & $0.61 $\\
$ 1024$ & $ 3.48$ & $0.62 $\\
\end{tabular}
\hskip 1cm\begin{tabular}{c|c|c}
$m$&$K_{\rm LL}$&$\ell_0$\\
\hline
$ \pm 1$ & $ 3.5$ & $ 0.63$ \\
$ \pm 2$ & $ 3.48$ & $0.6 $ \\
$ \pm 3$ & $3.5 $ & $ 0.63$ \\
$ \pm 4$ & $3.51 $ & $0.66 $ \\ 
$ \pm 5$ & $ 3.52$ & $0.68 $ \\
$ \pm 6$ & $ 3.54$ & $0.72$ \\
$ \pm 7$ & $ 3.65$ & $0.89 $ 
\end{tabular}
\caption{\label{tab:1}Parameters used to fit the QMC data $p_m(L/2)$ shown in Fig.~\ref{fig:qmc1} to the form Eq.~\eqref{eq:pmC}, either {\it vs.} $m$ at fixed $L$ (left) or {\it vs.} $L$ at fixed $m$ (right). The exact value of the Luttinger liquid parameter for  anisotropy $\Delta=-0.9$ is $K_{\rm LL}=3.4827$.}
\ec
\end{table}

\begin{figure}[h]
\centering
\includegraphics[width=.7\columnwidth,clip]{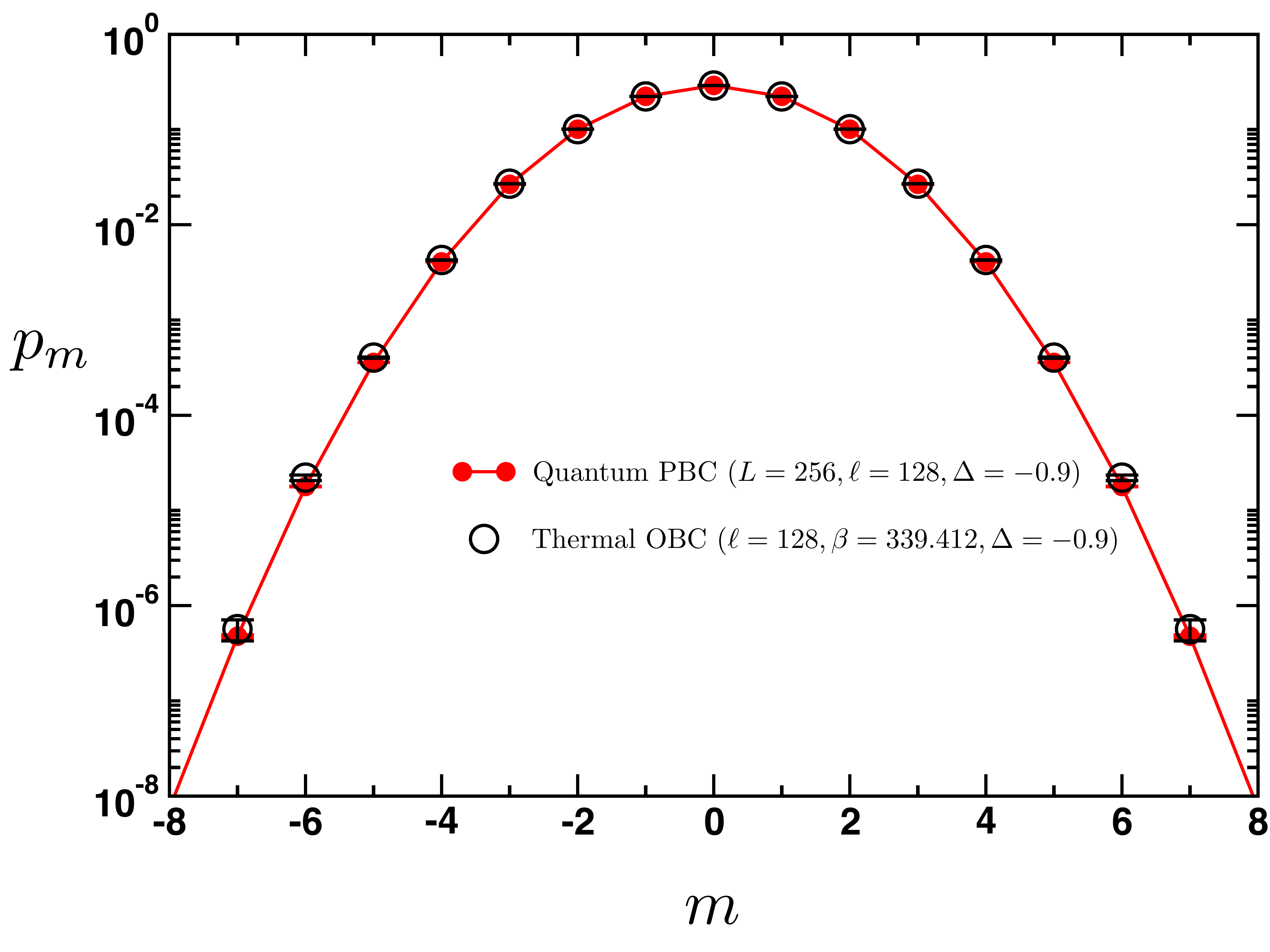}
\caption{QMC results for the weigths $p_m$ computed for XXZ chains with anisotropy $\Delta=-0.9$. ({\color{red}{$\bullet$}}) $L=256$, PBC and $T=0$ with a bipartition at $\ell=128$. ($\circ$) $\ell=128$, OBC, and $T=T_{\rm ent.}$ Eq.~\eqref{eq:Teff}.}
\label{fig:pm_q_th}
\end{figure}

\paragraph{Finite temperature---}
It is also instructive to test the validity of the quantum / thermal mapping Eq.~\eqref{eq:rhoA} by simply comparing $p_m$ for a bipartite system (XXZ with PBC) at $T=0$ with an open chain of length $\ell$ at finite temperature $T_{\rm ent.}=u\ln(\ell/\ell_0)/(\pi \ell)$. We take $L=256$ and $\ell=128$, with an anisotropy $\Delta=-0.9$, yielding $\ell_0=0.61$ from Tab.~\ref{tab:1}. Using the exact Bethe Ansatz expression for the velocity of excitations $u=\pi\sqrt{1-\Delta^2}/(2\arccos\Delta)$, we fix the entanglement temperature for this particular situation to $\beta_{\rm ent.}=339.412$, with no adjustable parameters. The comparison, shown in Fig.~\ref{fig:pm_q_th}, nicely validates the quantum/thermal correspondence.

\section{Conformal spectrum and spin resolved entanglement entropies}
Once the relative weights of spin-resolved sectors of the RDM  and the quantum/thermal mapping have been well understood, we now turn to the internal structure of the entanglement spectrum. Recently studied by L\"auchli in Ref.~\cite{lauchli_operator_2013}, we provide here further demonstration that it is directly related to the energy spectrum of an open XXZ chain. We then  discuss some consequences for the spin-resolved entanglement entropies.

\subsection{Conformal spectrum from DMRG}
Conformal field theory predicts \cite{alcaraz_surface_1987} the following low-energy spectrum for an open XXZ chain of $\ell$ sites
\be
E_0^m-E_0^0=\frac{\pi u}{2K_{\rm{LL}}\ell}m^2,
\label{eq:Em}
\ee
where $m$ is the $S^z$ quantum number, $E_0^0$ is the GS energy, $u$ the velocity of excitations, and $K_{\rm{LL}}$ the Luttinger liquid parameter. Such low-energy levels can be identified with the $q=\infty$ $m-$resolved entropies 
\be
S_{\infty}^{(m)}=-\ln\lambda_{\rm max}^{(m)},
\ee
provided the energy spectrum is correctly normalized. From the above definition of the RDM Eq.~\eqref{eq:rhoA}, and using the entanglement temperature Eq.~\eqref{eq:Teff}, the entropy is simply related to the above energy by $S=E/T_{\rm ent.}$, and therefore Eq.~\eqref{eq:Em} becomes
\be
S_{\infty}^{(m)}(\ell)-S_{\infty}(\ell)=\frac{{\cal B}\pi^2}{2K_{\rm{LL}}\ln(\ell/\ell_0)}m^2,
\label{eq:Sm}
\ee
where $S_{\infty}=-\ln\lambda_{\rm max}$ is the single copy entanglement~\cite{eisert_single-copy_2005}. 

\begin{figure}[t!]
\centering
\includegraphics[width=.85\columnwidth,clip]{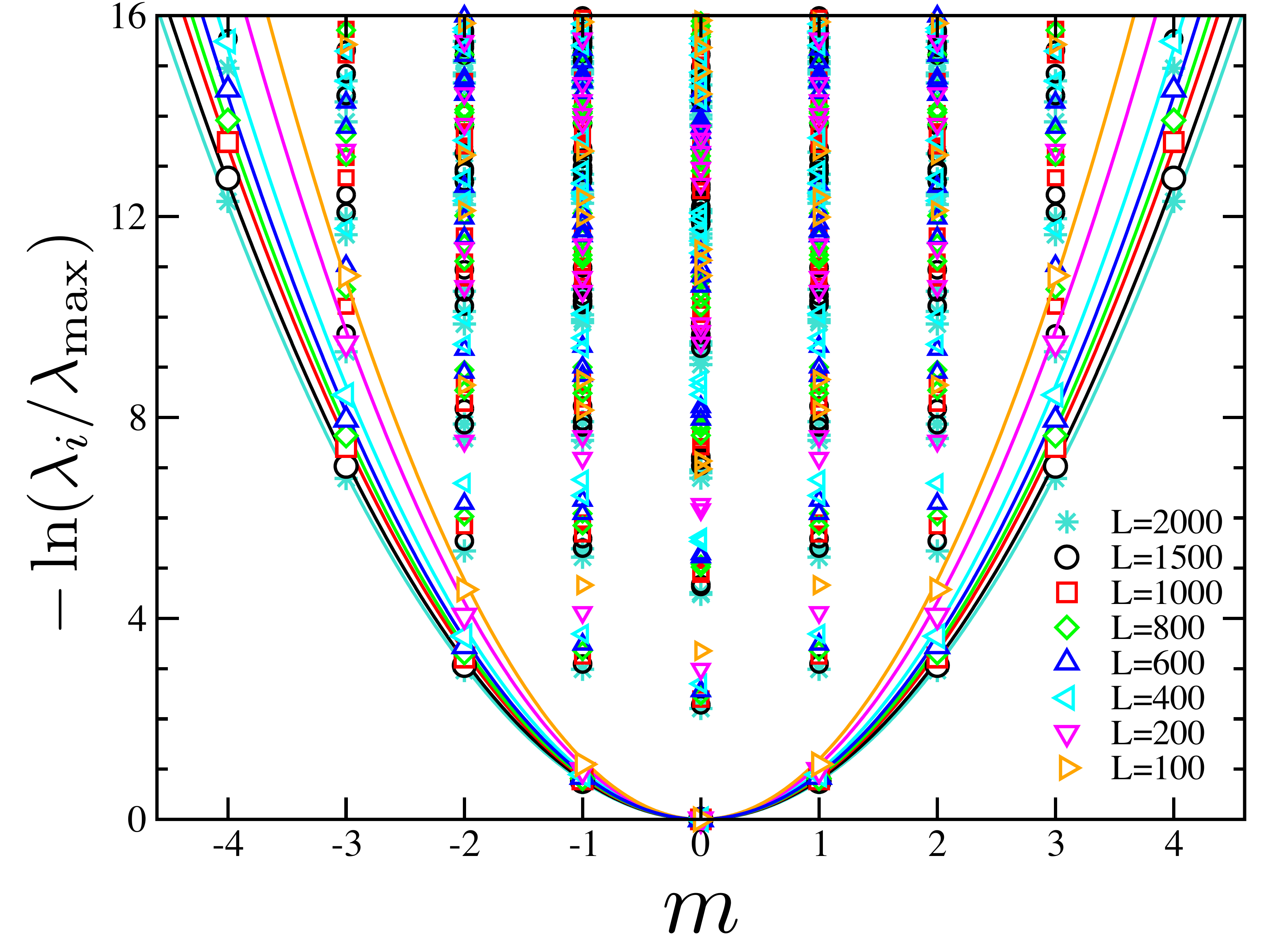}
\caption{Spin-resolved entanglement spectrum from DMRG calculations.  OBC results for XXZ chains at $\Delta=-0.5$ and various lengths $L$, as indicated on the plot. The lower part of the spectrum is fitted to the quadratic form Eq.~\eqref{eq:Sm}.}
\label{fig:ES_D-05}
\end{figure}

\begin{figure}[t!]
\centering
\includegraphics[width=.75\columnwidth,clip]{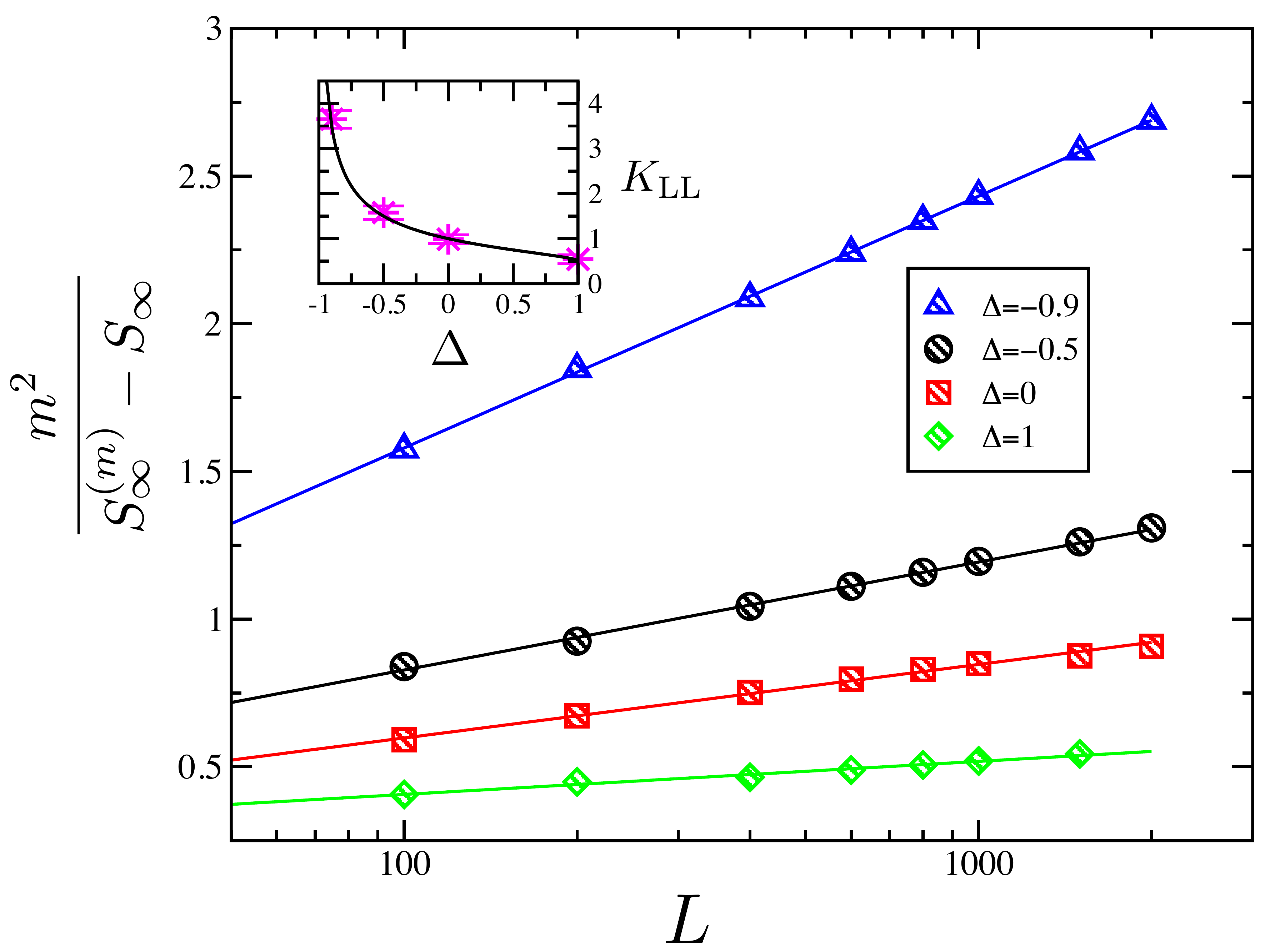}
\caption{DMRG results for the inverse curvature of the lower branch of the entanglement spectrum shown for various XXZ anisotropies {\it{vs.}} the chain length $L$. Solid lines are fits to the logarithmic scaling form Eq.~\eqref{eq:Sm} from which $K_{\rm LL}$ is extracted and showed in the inset against $\Delta$ (symbols) and compared to the exact expression Eq.~\eqref{eq:K} (black curve).}
\label{fig:prefactors}
\end{figure}

We have successfully checked this quadratic $m$ dependence using DMRG simulations for the XXZ model with OBC (setup (b) in Fig.~\ref{fig:OBC_PBC}), $\ell=L/2$, for various chain lengths $L=100,\ldots,2000$. Indeed, Fig.~\ref{fig:ES_D-05} shows the entanglement spectrum $-\ln(\lambda_i/\lambda_{\rm max})$ {\it{vs.}} the spin quantum number $m$ of subsystem $A$ for an Ising anisotropy $\Delta=-0.5$. The lower branch is precisely $S_{\infty}^{(m)}(\ell)-S_{\infty}(\ell)$, which fits perfectly to a parabola $\propto m^2$.
The prefactor of this quadratic form is studied in Fig.~\ref{fig:prefactors} where, plotted against $\ln L$, a very good agreement is found with Eq.~\eqref{eq:Sm}. This is also the case for other values of the Ising anisotropy $\Delta$.
According to Eq.~\eqref{eq:Sm} we expect the slope to be $K_{\rm{LL}}/\pi^2$, which compares very well to the exact result for the XXZ model Eq.~\eqref{eq:K}, as demonstrated in the inset of Fig.~\ref{fig:prefactors}.

At this stage, it is also interesting to remark that the curvature of the energy levels Eq.~\eqref{eq:Em} is controlled by the uniform susceptibility $\chi_0=\frac{K_{\rm{LL}}\ell}{u\pi}$ such as 
\be
E_0^m-E_0^0=\frac{m^2}{2\chi_0}.
\ee
Similarly, the quadratic entanglement spectrum
\be
S_{\infty}^{(m)}-S_{\infty}=\frac{m^2}{2C_2},
\ee
is controlled by the bipartite fluctuations of magnetization $C_2$.
\subsection{Consequences for the spin resolved entanglement entropies}
An important emerging question concerns the individual scalings of the spin-resolved 
von-Neumann entropies in each magnetization blocks $m$, defined by
\be
S_1^{(m)}=-\sum_i\lambda_i^{(m)}\ln\lambda_i^{(m)}.
\ee
The sum over the sectors $m$: $\sum_mS_1^{(m)}=S_1$ obeys the usual universal log scaling with the sub-system size $\ell$ Eq.~\eqref{eq:Sq}. It is therefore natural to ask whether some kind of universality may also emerge from invidual blocks, regarding their spin-resolved entanglement entropies.

\begin{figure}[t!]
\centering
\includegraphics[width=\columnwidth,clip]{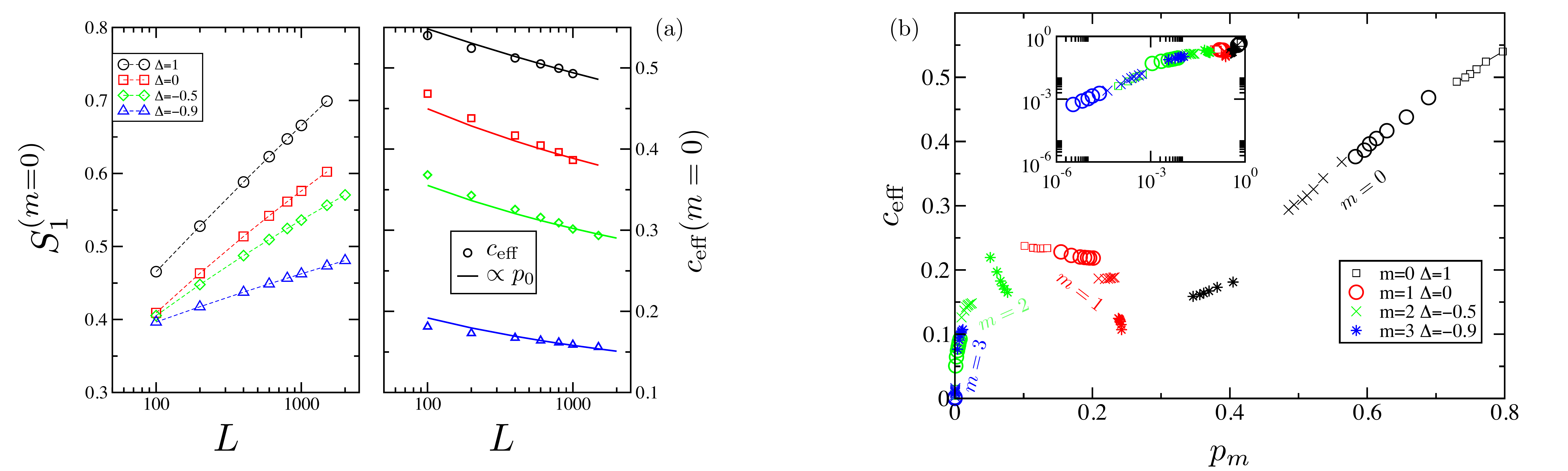}
\caption{(a) SREE $S_{1}^{(m=0)}$ plotted {\it{vs.}} the size $L$ for 4 values of the Ising anisotropy, as indicated on the plot. From the apparent log scaling of $S_{1}^{(m=0)}$, we extract the "effective central charge", defined in Eq.~\eqref{eq:ansatz}, plot it against $L$, and compare it to $\alpha_{\Delta}p_0(L)$ (full lines) with prefactors: $\alpha_1\simeq 0.66, \alpha_0\simeq 0.65, \alpha_{-0.5}\simeq 0.6, \alpha_{-0.9}\simeq 0.45$. (b) $c_{\rm eff}$ is plotted against $p_m$ for sectors $m=0, 1, 2, 3$ (different colors) and different anisotropies (different symbols). Inset: same in a log-log scale.}
\label{fig:SREE}
\end{figure}

For $q=\infty$ R\'enyi index, we have just seen that 
\bea
S_{\infty}^{(m)}(\ell)&=&S_{\infty}(\ell)+\frac{m^2}{2C_2(\ell)}\nonumber,\\
&=&\frac{c}{6{\cal B}}\ln (\ell/\ell_0)+\frac{{\cal B}\pi^2}{2K_{\rm{LL}}\ln(\ell/\ell_1)}m^2,
\label{eq:Smscaling}
\eea
where $\ell_0$ and $\ell_1$ are natural length scales of order 1.
From Eq.~\eqref{eq:Smscaling}, we see that for finite $m$ sectors, the leading scaling is $\propto c\ln \ell$ with additional slowly decaying corrections $\sim 1/\ln\ell$. 
However, consequences for the scaling of $S_1^{(m)}$ are not obvious. We can nevertheless try to make a conjecture. Using Jensen inequality\,\cite{jensen_sur_1906}, we have $H_q\ge H_{q'}$ if $q>q'$, where $H_q=(\ln\sum_i (x_i)^q)/(1-q)$ are normalized R\'enyi entropies, such that $\sum_ix_i=1$. This yields the following inequality for the spin-resolved entanglement entropies:
\be
{S_{1}^{(m)}}\ge p_m S_{\infty}^{(m)}.
\label{eq:ineq}
\ee
One can use the following ansatz for the SREE
\be
S_{1}^{(m)}(\ell)= \frac{c_{\rm eff}(m,\ell)}{3{\cal B}}\ln(\ell/\ell_0),
\label{eq:ansatz}
\ee
with \be \sum_mc_{\rm eff}(m,\ell)=1.\ee 
While there is no simple argument for the precise form of the "effective central charge" $c_{\rm eff}(m,\ell)$, the relation between $S_1$ and the single copy entanglement $S_\infty=S_1/2$ for critical chains leads us to make a conjecture, following Eq.~\eqref{eq:ineq}:
$$S_1^{(m)}(\ell)\stackrel{?}{=}2p_mS_{\infty}.$$
This would mean that the "effective central charge" $c_{\rm eff}(m,\ell)\approx p_m(\ell)$ slowly goes to zero with the system size. We have checked this conjecture against DMRG results, as shown in Fig.~\ref{fig:SREE} (a) for the $m=0$ sector and $\Delta=1, 0, -0.5, -0.9$. The left panel of Fig.~\ref{fig:SREE} (a) shows $S_1^{(m=0)}(L)$ in a log-linear scale from which, according to the ansatz Eq.~\eqref{eq:ansatz}, the "effective central charge" $c_{\rm eff} (m=0,L)$ is extracted and plotted in the right panel of Fig.~\ref{fig:SREE} (a). We find that $c_{\rm eff} (m=0,L)$ is slowly decaying with $L$, in qualitative agreement with the decay of $p_0(L)$ [Eq.~\eqref{eq:pm}]. Indeed, the values of $c_{\rm eff}(L)$ compare relatively well to $\alpha_\Delta p_0(L)$ with prefactors $\alpha_1\simeq 0.66, \alpha_0\simeq 0.65, \alpha_{-0.5}\simeq 0.6, \alpha_{-0.9}\simeq 0.45$. However, this scaling becomes less good for the other sectors $m\neq 0$, as visible in panel (b) of Fig.~\ref{fig:SREE} where the linear behavior $c_{\rm eff}(m,\ell)\sim p_m(\ell)$ does not appear to be valid, at least for the sizes considered here.

Therefore, based on our finite size data we cannot conclude on the validity of the proposed log scaling Eq.~\eqref{eq:ansatz}, and regarding the putative "effective central charge" $c_{\rm eff}(m,\ell)$. Further studies are necessary in order to capture the scaling features of the spin-resolved entanglement entropies.

\section{Conclusions}
In order to elucidate the structure, properties, and meaning of the real space entanglement spectrum we have investigated the XXZ spin 1/2 chain in the critical Luttinger liquid regime. Particular emphasis has been brought to the presence of the additional spin quantum number associated with spin conservation. We have further elaborated on the quantum/thermal correspondence between the entanglement spectrum of a $T=0$ pure quantum state and the thermal density matrix of an effective entanglement Hamiltonian at a finite entanglement temperature $T_{\rm ent}\sim \ln(\ell)/\ell$. This allowed us to identify a direct correspondence between the entanglement spectrum of an XXZ chain with the energy spectrum of an open XXZ chain. In the second part of the paper, we have introduced entanglement entropies for each individual block of the reduced density matrix associated with the spin quantum number of the subsystem, dubbed spin-resolved entanglement entropies. We proposed the scaling behavior with the subsystem length of these new entropies.
Within the process of exploring the spin-resolved properties of the reduced density matrix and the entanglement spectrum in general, various interesting questions have arisen which remain to be clarified:

(i) In Fig.2 we have shown the eigenvalue distribution of the reduced density matrix for the XXZ chain and compared to the analytical prediction of Calabrese and Lefevre~\cite{calabrese-08pra032329}. While the free fermion case $\Delta=0$ agrees very well, interacting cases ($\Delta\neq 0$) show a significant deviation from the Calabrese-Lefevre result. Finite size extrapolation revealed that the leading correction is $\sim 1/\ln L$. Surprisingly, this correction seems to change its sign with the sign of the Ising anisotropy $\Delta$. At the free fermion point, we did not find any log-correction suggesting the correction to be of the form $\sim \Delta/\ln L$. Analytical understanding of such a finite-size correction is desirable.

(ii) For the introduced spin-resolved entanglement entropies, we made the conjecture $S^{(m)}_{1}(\ell)=2p_mS_{\infty}= c_{\rm eff}(m,\ell)/(3{\cal B}) \ln (\ell/\ell_0)$ implying that the spin-resolved entanglement entropies are controlled by an effective central charge $c_{\rm eff}(m,\ell)\sim 1/\sqrt{\ln(\ell/\ell_1)}$ which slowly goes to zero with the system size. Although we were using very large-scale DMRG and QMC numerical simulations, we could not draw a firm conclusion regarding this point. A better analytical understanding would be needed, probably within the framework of conformal field theory. A possible calculation would be to compute the low temperature behavior of the thermal entropy of the entanglement Hamiltonian within each magnetization sector.

Finally, it would be very interesting to extend these ideas of spin-resolved entanglement spectra and entropies to other strongly correlated systems, as well as to higher dimensional systems~\cite{alba_2013,luitz_shannon-renyi_2014,luitz_participation_2014}.\\

\noindent {\bf Acknowledgements}

It is a pleasure to thank P. Calabrese, F. Pollmann, J. Dubail for interesting conversations related to this work and P. Schmitteckert for supporting us with the data for chains with $L=2000$. NL is supported by the French ANR program ANR-11-IS04-005-01. SR is supported by the DFG through FOR 960 and by the Helmholtz association through VI-521

\appendix
\section{DMRG convergence}
\label{sec:appendix}

As exemplified below in Fig.~\ref{fig:comp} for a $L=2000$ open XX chain, when keeping only eigenvalues $\lambda>10^{-9}$, the von Neumann entropy of  a half chain $S_1=-\sum_{i=1}^{N_{\lambda}}\lambda_i\ln\lambda_i$ (with $\lambda_1>\lambda_2>\cdots$) converges very rapidly with the number $N_\lambda$. In Fig.\,\ref{fig:comp} we observe that the lowest 20 entanglement levels are sufficient to reach the exact value of $S_1$ within $1\%$ accuracy. In the inset of Fig.~\ref{fig:comp} the difference between the exact result and the DMRG data as a function of $N_\lambda$ is shown, further substantiating the good convergence behavior.

\begin{figure}[h!]
\centering
\includegraphics[width=.5\columnwidth,clip]{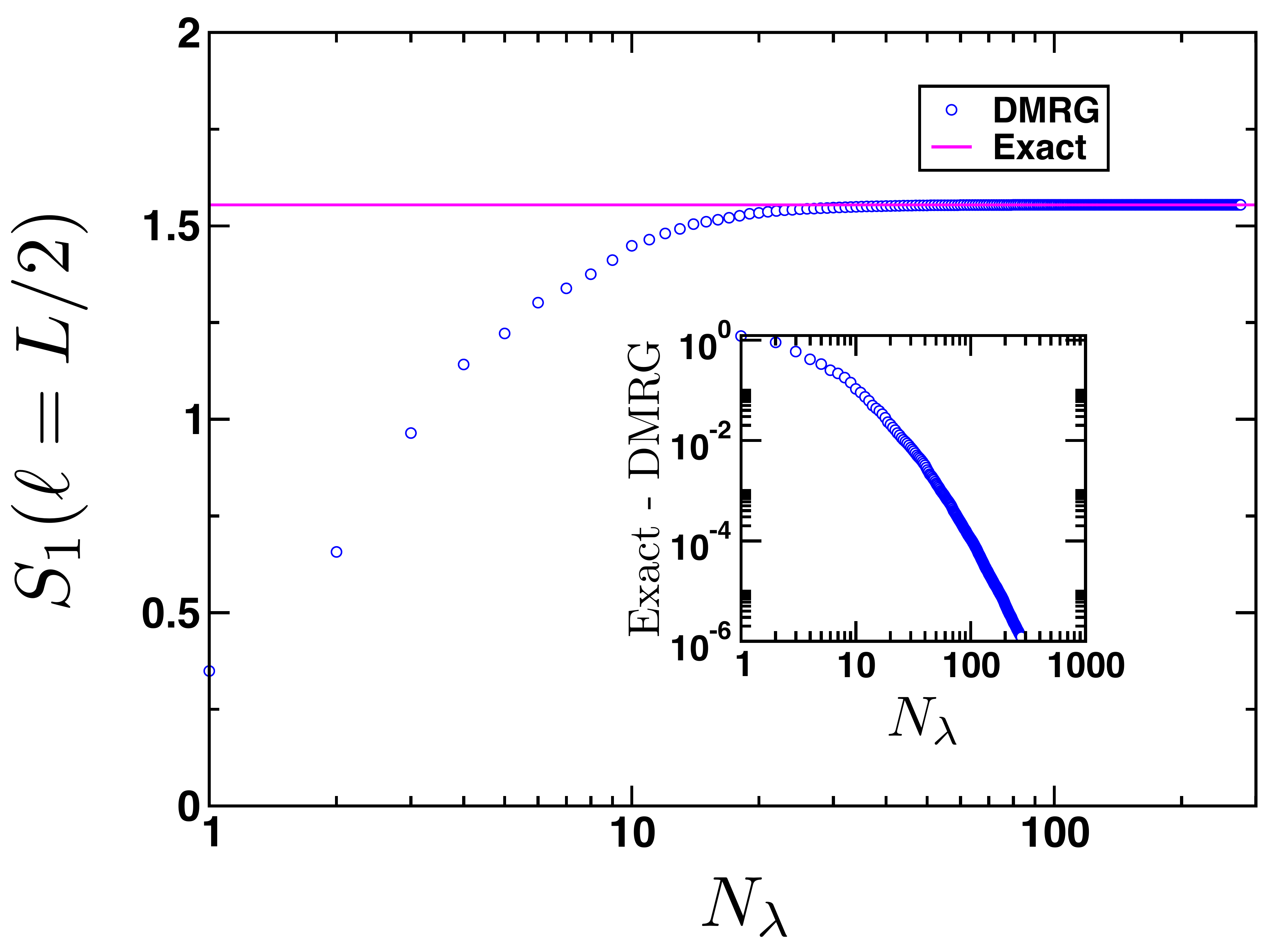}
\caption{Comparison between DMRG and ED for $S_1(\ell=L/2)$ of an open XX chain of length $L=2000$ sites.}
\label{fig:comp}
\end{figure}
\newpage
\bibliographystyle{modernref}

\bibliography{jstat}

\end{document}